\documentclass[aps,prc,showpacs,twocolumn,epsfig]{revtex4}
\usepackage[dvips]{graphicx}
\begin{document}

\draft
\title{Surface tension in a compressible liquid-drop model: Effects on nuclear
density and neutron skin thickness}

\author{Kei Iida$^1$ and Kazuhiro Oyamatsu$^{1,2,3}$}
\affiliation{$^1$The Institute of Physical and Chemical Research (RIKEN),
Hirosawa, Wako, Saitama 351-0198, Japan\\
$^2$Department of Media Theories and Production, Aichi Shukutoku
University, Nagakute, Nagakute-cho, Aichi-gun, Aichi 480-1197, Japan\\
$^3$Department of Physics, Nagoya University, Furo-cho, Chigusa-ku, 
Nagoya, Aichi 464-8602, Japan}

\date{\today}

\begin{abstract}

     We examine whether or not the surface tension acts to increase the 
nucleon density in the nuclear interior within a compressible liquid-drop 
model.  We find that it depends on the density dependence of the surface
tension, which may in turn be deduced from the neutron skin thickness of
stable nuclei.

\end{abstract}
\maketitle

      The saturation property of bulk nuclear matter is usually deduced from 
empirical data on the masses and radii of stable nuclei \cite{BW}.  If nuclear 
matter is incompressible, the bulk saturation density is equal to the density 
in the nuclear interior.  However, nuclear matter in nuclei is more or less
compressible.  The question of how the finite compressibility of nuclear 
matter makes the density in the nuclear interior deviate from the saturation 
density has yet to be clarified.  In this paper, by using a compressible 
liquid-drop model, we show that a key feature in addressing this question is 
the density dependence of the surface tension.  We then derive its relation 
with the neutron skin thickness.

      The dependence of the surface energy on the density of the nucleon
liquid naturally occurs.  In the Fermi-gas model, the surface energy arises
from a reduction of the available density of states of nucleons due to the
presence of the surface, beyond which the nucleon flux vanishes.  The total
kinetic energy thus increases by an amount proportional to the surface area,
and this increase depends on the nucleon density through the nucleon Fermi 
momentum.  In addition to this kinetic contribution, the interaction between 
nucleons contributes to the surface energy more importantly.  This is because a
nucleon in the surface region does not perceive the same amount of attraction
from the surrounding nucleons as that which it would if it were in the deeper 
region.  This contribution is determined by the property of the nuclear 
medium, which is characterized by the nucleon density.  We remark that for the
same reasons, the surface energy depends also on the neutron excess of the 
liquid; this dependence has to be invariant under exchange between neutrons 
and protons.

      As discussed by Yamada \cite{Ya}, the dependence of the surface energy
on the inner liquid density controls the density deviation from the bulk 
saturation density.  Usually, this dependence is not considered explicitly. 
This is because the surface tension is normally calculated for a planar 
interface between the saturated nucleon liquid and the vacuum, which are in 
mechanical equilibrium.  However, this is not equivalent to mechanical 
equilibrium in a real nucleus, which is generally associated with additional 
pressures arising from the size- and density-dependence of the surface energy.
These pressures in turn affect the equilibrium density of the compressible 
nucleon liquid. 

      In order to see this effect, we utilize a compressible liquid-drop 
model, which gives rise to a semi-empirical mass formula in a way dependent on
the density and neutron excess in the nuclear interior, $n_{\rm in}$ and 
$\delta_{\rm in}$.  Generally, a liquid-drop model is advantageous to the 
description of various macroscopic properties of nuclei.  During the past 
decade, it has been used in describing, e.g., neutron skin \cite{Da},
nuclear fission \cite{Nat}, deformation of rapidly rotating nuclei \cite{MS3},
synthesis of superheavy nuclei \cite{MS3}, and nuclei in neutron star crusts 
\cite{PR2}.

      Throughout this paper we consider nearly symmetric nuclei,
for which we can set $R_n\simeq R_p$.  We assume that the distribution of $i$ 
nucleons ($i=n,p$) is spherically symmetric, uniform at a number density 
$n_i$, and squared off at a radius $R_i$.  For a nucleus of mass number $A$ 
and charge number $Z$ (neutron number $N=A-Z$), we thus obtain 
$n_{\rm in}=n_n+n_p$ and
\begin{eqnarray}
\delta_{\rm in} &=& (n_n-n_p)/(n_n+n_p)
 \nonumber \\  &\simeq& (N-Z)/A.
\end{eqnarray}
We then write the binding energy $E_B$ of the nucleus as
\begin{equation}
-E_B=E_{\rm vol} + E_{\rm surf} +E_{\rm Coul}.
\label{bind}
\end{equation}
Here 
\begin{equation}
E_{\rm vol}=A w(n_{\rm in},\delta_{\rm in}),
\label{Evol}
\end{equation}
with the bulk energy per nucleon $w$, is the volume energy,
\begin{equation}
E_{\rm surf}=4\pi\sigma(n_{\rm in},\delta_{\rm in})R_p^2,
\label{Esurf}
\end{equation}
with the density-dependent surface tension $\sigma$, is the surface energy, 
and
\begin{equation}
E_{\rm Coul}=\frac{3 Z^2 e^2}{5R_p},
\label{Ecoul}
\end{equation}
is the Coulomb energy.  For $w$ and $\sigma$, we adopt a form expanded with 
respect to the density and neutron excess around $n_{\rm in}=n_0$ and 
$\delta_{\rm in}=0$:
\begin{eqnarray}
w(n_{\rm in},\delta_{\rm in})&=&w_0+\frac{K_0}{18n_0^2}(n_{\rm in}-n_0)^2
    \nonumber \\ & &
      +\left[S_0+\frac{L}{3n_0}(n_{\rm in}-n_0)\right]\delta_{\rm in}^2,
\label{bulk}
\end{eqnarray}
where $n_0$ and $w_0$ are the saturation density and energy of symmetric 
nuclear matter, $K_0$ is the incompressibility of symmetric nuclear matter, 
$S_0$ is the symmetry energy coefficient, and $L$ is the density symmetry 
coefficient, and 
\begin{equation}
\sigma(n_{\rm in},\delta_{\rm in})=\sigma_0\left[1-C_{\rm sym}\delta_{\rm in}^2
                 +\chi\left(\frac{n_{\rm in}-n_0}{n_0}\right)\right],
\label{sigma}
\end{equation}
where $\sigma_0=\sigma(n_0,0)$, $C_{\rm sym}$ is the surface symmetry energy
coefficient, and $\chi=(n_0/\sigma_0)\partial \sigma / \partial 
n_{\rm in}|_{n_{\rm in}=n_0,\delta_{\rm in}=0}$.  In Eq.\ (\ref{bind}) we have
ignored the energy contribution of the neutron skin thickness $R_n-R_p$, which
will be considered later, and curvature corrections.  We have also ignored 
pairing and shell corrections since we will confine ourselves to macroscopic 
properties of the nuclear ground state.  We remark that in equilibrium 
$n_{\rm in}$ is related to $\delta_{\rm in}$, as we shall see just below.
 
    Some of the coefficients characterizing the bulk energy (\ref{bulk}) can 
be deduced from empirical data for the masses and root-mean-square charge 
radii of stable nuclei.  The saturation density $n_0$, the saturation energy
$w_0$, and the symmetry energy coefficient $S_0$ typically take on a value 
ranging 0.14--0.17 fm$^{-3}$, $-16\pm1$ MeV, and 25--40 MeV, whereas the 
incompressibility $K_0$ and the density symmetry coefficient $L$, which 
control the density dependence of bulk nuclear matter, are not well 
constrained.  Using a simplified version of the Thomas-Fermi model \cite{OI} 
we found that various sets of the values of $K_0$ and $L$ ranging 180--360 MeV
and 0--200 MeV reasonably reproduce the empirical masses and radii and that 
future systematic measurements of the matter radii of unstable neutron-rich 
nuclei would give a good constraint on the value of $L$.

    We turn to the coefficients in the surface tension (\ref{sigma}).  The
primary coefficient $\sigma_0$ and the surface symmetry coefficient 
$C_{\rm sym}$ can be estimated from the empirical mass data as 
$\sigma_0\simeq 1$ MeV fm$^{-2}$ and $C_{\rm sym}=1.5$--2.5.  The parameter
$\chi$ characterizing the density dependence of the surface tension is poorly 
known and hence the quantity of interest in this work.  Myers and Swiatecki 
\cite{MS} simply set $\chi=0$, while the Fermi-gas model predicts $\chi=4/3$.  
We will see that a precise determination of $n_0$ from the values of 
$n_{\rm in}$ deduced, e.g., from electron-nucleus elastic scattering data
requires reliable information about $\chi$.

    Let us now estimate the equilibrium value of $n_{\rm in}$ from pressure
equilibrium and compare it to the bulk saturation density $n_s$ at fixed 
$\delta_{\rm in}$.  Within the present compressible liquid-drop model, the 
pressure equilibrium condition can be obtained from optimization of the 
binding energy (\ref{bind}) with respect to the size under fixed $A$ and $Z$ as
\begin{equation}
 0=P_{\rm vol} + P_{\rm surf} +P_{\rm Coul}.
\label{peq}
\end{equation}
Here 
\begin{eqnarray}
P_{\rm vol}&=&\frac{K_0}{9}(n_{\rm in}-n_0)+\frac{L}{3}n_0 \delta_{\rm in}^2
  \nonumber  \\ &\equiv&\frac{K_0}{9}(n_{\rm in}-n_s)
\label{Pvol}
\end{eqnarray}
is the volume pressure,
\begin{equation}
P_{\rm surf}=-\frac{2\sigma_0}{R_p}\left[1-\frac{3}{2}\chi
              -C_{\rm sym}\delta_{\rm in}^2
              +\chi\left(\frac{n_{\rm in}-n_0}{n_0}\right)\right]
\label{Psurf}
\end{equation}
is the surface pressure, and
\begin{equation}
P_{\rm Coul}=\frac{3 Z^2 e^2}{20 \pi R_p^4},
\label{Pcoul}
\end{equation}
is the Coulomb pressure.  The bulk pressure vanishes at the saturation density,
\begin{equation}
n_s=n_0-\frac{3Ln_0}{K_0}\delta_{\rm in}^2,
\label{sd}
\end{equation}
which generally decreases with increasing $\delta_{\rm in}$ as has already 
been discussed in Refs.\ \cite{OI,OTSST}.  The Coulomb pressure acts to 
increase the nuclear size, whereas the surface pressure tends to enlarge or
reduce the nuclear size according to whether $\chi$ is larger or smaller than
$\sim2/3$.

    The relation (\ref{peq}), if the Coulomb pressure $P_{\rm Coul}$ is 
ignored, can be reduced to Laplace's formula.  This can be done by 
transforming Eq.\ (\ref{peq}) into
\begin{equation}
  P_{\rm vol}+\frac{3\sigma_0\chi}{R_p}
  =\frac{2\sigma(n_{\rm in},\delta_{\rm in})}{R_p}.
\end{equation}
Here the left side, arising from the energy derivative with respect to 
$n_{\rm in}$, corresponds to the pressure of the nucleon liquid, 
while the right side arises from the energy derivative with respect to $R_p$.

     Deviation of the equilibrium value of $n_{\rm in}$ from the bulk 
saturation density $n_s$ at fixed $\delta_{\rm in}$ can be estimated from 
condition (\ref{peq}) as
\begin{eqnarray}
n_{\rm in}-n_s&\simeq&0.016 \left(\frac{230~{\rm MeV}}{K_0}\right)
                      \left(\frac{\sigma_0}{1~{\rm MeV~fm}^{-2}}\right)
                      \left(\frac{5~{\rm fm}}{R_p}\right)
   \nonumber  \\ & & \times \left(1-\frac{3}{2}\chi 
      -\frac{3 Z^2 e^2}{40 \pi R_p^3 \sigma_0}\right)~{\rm fm}^{-3}.
\label{neq}
\end{eqnarray}
In this estimate we have used $P_{\rm surf}\simeq-2\sigma_0(1-3\chi/2)/R_p$.
The ratio of the Coulomb pressure to the surface pressure, $3 Z^2 e^2/40 \pi 
R_p^3 \sigma_0$, is typically 0.2--0.6.  We find from this estimate that for 
$\chi$=0 and 4/3, the surface pressure can induce about 10 \% change in 
$n_{\rm in}$ in different directions.  We thus see the role played by $\chi$ 
in determining $n_s$ from empirical information about $n_{\rm in}$.

     We now proceed to show that the neutron skin thickness, which has been
neglected so far, is a quantity that may be useful for deduction of the value
of $\chi$.  In doing so, as considered by Pethick and Ravenhall \cite{PR}, it 
is convenient to describe the nuclear surface in a thermodynamically consistent
manner.  In this description, the nuclear surface is in thermodynamic 
equilibrium with the bulk system composed of $A$ nucleons, and the neutron
skin arises from adsorption of $N_s$ neutrons onto the nuclear surface.  
The interior region composed of $A-N_s$ nucleons acts as a reservoir of
neutron chemical potential $\mu_n$, and neutrons can go back and forth between
the skin and interior regions.  Consequently, the relevant thermodynamic 
quantity is the thermodynamic potential, $\Omega=\Omega_{\rm vol} + 
\Omega_{\rm surf} +\Omega_{\rm Coul}$, divided in a similar way to the
binding energy (\ref{bind}).  In equilibrium, $\Omega_{\rm surf}=\sigma 
{\cal A}$, where ${\cal A}$ is the surface area.  A small quasistatic change 
in the neutron excess in the interior region with $A$ and $N$ fixed gives rise
to a change in the thermodynamic potential of the surface, $\Delta
\Omega_{\rm surf}$, and a change in the neutron chemical potential, $\Delta 
\mu_n$, which are related as
\begin{equation}
   \Delta\Omega_{\rm surf}
   = - N_s \Delta \mu_n.
\label{theq}
\end{equation}
This relation indicates that the neutron skin can be described in terms of
the bulk and surface properties.  We remark that at $N=Z$, a balance between 
the induced changes $\Delta\Omega_{\rm vol}$ and $\Delta\Omega_{\rm Coul}$ in 
the volume and Coulomb energies allows the neutron excess in the interior 
region to deviate from zero and hence a proton skin to occur, as we shall see.

     It is straightforward to combine the above thermodynamic description of
the nuclear surface with the compressible liquid-drop model adopted here.
In this model, $\delta_{\rm in}$, $N_s$, ${\cal A}$, and $\mu_n$ read 
\begin{eqnarray}
\delta_{\rm in} &=& \frac{N-N_s-Z}{A-N_s}
  \nonumber \\ &\simeq& 
\frac{N-Z}{A}-\frac{3(R_n-R_p)}{2R_p},
\label{deltain}
\end{eqnarray}
\begin{equation}
N_s \simeq 4\pi R_p^2 n_n (R_n-R_p), ~~~{\cal A} \simeq 4\pi R_p^2,
\label{Ns}
\end{equation}
and
\begin{equation}
\mu_n=w_0+S_0\delta_{\rm in}(2-\delta_{\rm in})+{\cal O}(\delta_{\rm in}^3).
\label{mun}
\end{equation}
Here we have calculated $\mu_n$ at the bulk saturation density (\ref{sd});
the dependence of $\mu_n$ on the parameters $L$ and $K_0$ characterizing the 
density dependence of the bulk energy does not appear up to second order in 
$\delta_{\rm in}$.

    We can now obtain the expression for the neutron skin thickness in the 
absence of Coulomb energy.  In this case the system is symmetric under 
exchange between neutrons and protons.  Substitution of Eqs.\ (\ref{sigma}), 
(\ref{deltain}), (\ref{Ns}), and (\ref{mun}) into Eq.\ (\ref{theq}) leads to 
\begin{equation}
   R_n-R_p=C\delta\left(1+\frac{3C}{2R_p}\right)^{-1}+{\cal O}(\delta^2),
\label{thick0}
\end{equation}
where 
\begin{equation}
 C \equiv \frac{2\sigma_0}{S_0 n_0}\left(C_{\rm sym}+\frac{3L\chi}{K_0}\right),
\label{C}
\end{equation}
and $\delta\equiv(N-Z)/A$.  The parameter $C$ originates mainly from a change 
in the surface tension due to the small quasistatic change in 
$\delta_{\rm in}$.  This change is characterized not only by the surface 
symmetry energy coefficient $C_{\rm sym}$, but also by the parameter $\chi$ 
through the dependence of the saturation density $n_s$ given by Eq.\ (\ref{sd})
on $\delta_{\rm in}$.  The term $3L\chi/K_0$ in the right side of Eq.\ 
(\ref{C}) is associated with the finite compressibility and thus vanishes in 
the incompressible limit in which Eq.\ (\ref{thick0}) reduces to the result for
$R_n-R_p$ obtained by Pethick and Ravenhall \cite{PR}.  Note that this term 
does not exist in the result of Myers and Swiatecki \cite{MS2} who presumed 
$\chi=0$ in a compressible liquid-drop picture, although it can be comparable 
with $C_{\rm sym}$.  The fact that $3L\chi/K_0$ is poorly known suggests that 
one could not deduce the EOS parameters $L$ and $K_0$ from experimental data 
for the neutron skin thickness without knowing $\chi$.  Consequently, it turns
out that previous investigations that attempted to relate the neutron skin 
thickness with the EOS of nuclear matter \cite{OTSST,SL,Da} do not take full 
account of uncertainties in the parameters $\chi$, $L$, and $K_0$.
We can see from Eq.\ (\ref{thick0}) that the neutron skin vanishes at $N=Z$, 
as it should in the absence of Coulomb energy.

     Coulomb effects ignored in Eq.\ (\ref{thick0}) induce a thin proton skin 
at $N=Z$ through a deviation of $\delta_{\rm in}$ from zero and a polarization
of the nuclear interior, as discussed by Myers and Swiatecki \cite{MS2}. 
First, in order to calculate the deviation of $\delta_{\rm in}$ at $N=Z$, one 
has only to consider the balance between $\Delta\Omega_{\rm vol}$ and 
$\Delta\Omega_{\rm Coul}$.  In this case an increment in the proton radius 
$R_p$ tends to reduce the Coulomb energy, while it leads to a cost of the
symmetry energy in the nuclear interior.  We may thus obtain
\begin{equation}
  \delta_{\rm in} \simeq \frac{Ze^2}{20 R_p S_0} ~~~\mbox{at $N=Z$},
\label{delta0}
\end{equation}
which typically amounts to a small value of $\sim0.02$.  For the neutron skin 
thickness (\ref{thick0}), this deviation effectively replaces $\delta$ by 
$\delta-Ze^2/20R_p S_0$ \cite{MS2}.  In the liquid-drop picture, the $\beta$ 
stability condition that the neutron and proton chemical potentials are almost
equal is given by
\begin{equation}
  \delta_{\rm in} \simeq \frac{3Ze^2}{10 R_p S_0}.
\label{deltabeta}
\end{equation}
This $\delta_{\rm in}$ is much larger than $Ze^2/20R_p S_0$, which implies 
that generally stable nuclei, lying around the $\beta$ stability line, have a 
neutron skin.  Second, the polarization of the nuclear interior tends to 
reduce a difference between the root-mean-square radii of the neutron and 
proton density distributions, $\Delta r_{np}$, by redistributing nucleons in 
such a way as to deplete protons in the central region.  This reduction 
amounts to $\sqrt{3/5}Ze^2/70S_0$ \cite{MS2}, which typically takes on a small
value of 0.01--0.04 fm.

     By incorporating these Coulomb effects into Eq.\ (\ref{thick0}), we 
finally obtain
\begin{equation}
   \Delta r_{np} \simeq \sqrt{\frac{3}{5}}
            \left[C\left(\delta-\frac{Ze^2}{20R_p S_0}\right)
             \left(1+\frac{3C}{2R_p}\right)^{-1} - \frac{Ze^2}{70S_0}\right],
   \label{thick}
\end{equation}
where a factor $\sqrt{3/5}$ arises from a difference between the 
root-mean-square and the half-density radius for the rectangular distribution.
Here we have ignored corrections due to the difference in the surface 
diffuseness between protons and neutrons.  These corrections are lower order 
in $A$ than the terms in Eq.\ (\ref{thick}) \cite{Da} and hence expected to be
negligibly small for heavy stable nuclei of interest here.

      We now ask how one can obtain information about $\chi$ from empirically
deduced values of $\Delta r_{np}$ for stable nuclei such as Ni and Sn isotopes.
Such values can be deduced, e.g., from measurements of proton- and 
electron-nucleus elastic differential cross sections \cite{Ba}.  It is 
instructive to compare the deduced values with the prediction by the present 
liquid-drop model, which is given by Eq.\ (\ref{thick}). 
Figure 1 exhibits the deduced and predicted values as a function of $\delta$. 
When $\chi=0$, the predicted values by setting the other parameters at typical
values are appreciably smaller than the deduced ones.  This suggests that 
$\chi$ is likely to be positive.  However, the magnitude of $\chi$ remains to 
be clarified since $\chi$ is coupled with the uncertain parameters $L$ and 
$K_0$ in Eq.\ (\ref{thick}).  We remark that a staggering of the deduced values
is far larger than that of the predicted values.  This may be partly because
the former values were deduced by various groups using different models for 
proton elastic scattering and nucleon density distribution, and partly because
pairing and shell effects are ignored in the present prediction.

\begin{figure}[t]
\begin{center}
\includegraphics[width=8.5cm]{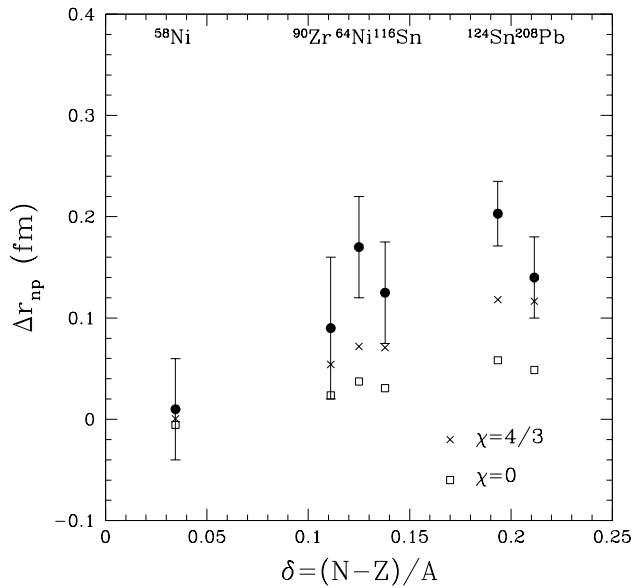}
\end{center}
\vspace{-0.5cm}
\caption{\label{skin}
Difference in the root-mean-square radius between neutrons and 
protons for six stable nuclei of $A>50$.  The squares and crosses denote the 
results calculated from Eq.\ (\ref{thick}) for $\chi=0$ and 4/3; the other 
parameters are set to be $n_0=0.16$ fm$^{-3}$, $S_0=30$ MeV, $K_0=230$ MeV, 
$L=100$ MeV, $\sigma_0=1$ MeV fm$^{-2}$, $C_{\rm sym}=1.8$, and 
$R_p=1.2A^{1/3}$ fm.  The empirical data (dots) are taken from Refs.\ 
\cite{Ba,Te}.  
}
\end{figure}

     In summary we have found from a compressible liquid-drop model that 
whether or not the nucleon density in the nuclear interior is larger than the 
bulk saturation density depends on the density dependence of the surface 
tension, which in turn controls the neutron excess dependence of the neutron 
skin thickness.  In order to deduce the density dependence of the surface 
tension and the bulk saturation density from the neutron skin thickness and the
interior density, it would be useful to systematically analyze differential 
cross sections measured for proton and electron elastic scattering off stable 
nuclei.  In such analysis of proton elastic scattering data obtained for 
incident energies above 500 MeV, one could relate the angle of diffraction 
maxima measured in the small momentum transfer region to the 
root-mean-square matter radius by using the Glauber theory in the optical 
limit approximation \cite{IOB}.  The research in this direction is under way.

      We are grateful to A. Kohama for helpful discussions.  This work was 
supported in part by RIKEN Special Postdoctoral Researchers Grant No.\ 
A11-52040.

\end{document}